\documentclass[sigconf]{acmart}

\AtBeginDocument{%
  }

\copyrightyear{2026}
\acmYear{2026}
\setcopyright{cc}
\setcctype{by}
\acmConference[CIKM '26] {Proceedings of the 35th ACM International Conference on Information and Knowledge Management}{November 7--11, 2026}{Rome, Italy.}
\acmBooktitle{Proceedings of the 35th ACM International Conference on Information and Knowledge Management (CIKM '26), November 7--11, 2026, Rome, Italy}
\acmISBN{979-8-4007-2539-5/2026/11}
\acmDOI{10.1145/3799682.3840569}

\newcommand{\std}[1]{{\textcolor{gray!100}{±#1}}}

\usepackage{enumitem}
\usepackage{colortbl}
\usepackage{multirow}
\usepackage{algorithm}
\usepackage{algpseudocode}
\usepackage{microtype}
\usepackage{balance}
\usepackage{float}
\begin{document}

\title{Code-as-Auditor: Executable Compliance Reasoning via Regulation-to-Code}


\author{Jisoo Kim}
\orcid{0009-0001-8782-1279} 
\affiliation{%
  \institution{Sungkyunkwan University}
  \city{Suwon}
  \country{Republic of Korea}
}
\email{clrdln@g.skku.edu}

\author{Taeyoon Kwack}
\orcid{0009-0008-7299-9534} 
\affiliation{%
  \institution{Sungkyunkwan University}
  \city{Suwon}
  \country{Republic of Korea}
}
\email{njj05043@g.skku.edu}

\author{Jinwoo Jang}
\orcid{0009-0003-8943-3565} 
\affiliation{%
  \institution{Sungkyunkwan University}
  \city{Suwon}
  \country{Republic of Korea}
}
\email{jinustar@g.skku.edu}

\author{Woo Kyung Kim}
\orcid{0000-0001-6214-4171} 
\affiliation{%
  \institution{Sungkyunkwan University}
  \city{Suwon}
  \country{Republic of Korea}
}
\email{kwk2696@g.skku.edu}

\author{Honguk Woo}
\orcid{0000-0001-6948-3440} 
\affiliation{%
  \institution{Sungkyunkwan University}
  \city{Suwon}
  \country{Republic of Korea}
}
\authornote{Corresponding author.}
\email{hwoo@skku.edu}

\renewcommand{\shortauthors}{Jisoo Kim, Taeyoon Kwack, Jinwoo Jang,Woo Kyung Kim, \& Honguk Woo}

\begin{abstract}
  Large Language Models (LLMs) are increasingly adopted for compliance and legal reasoning tasks, yet their outputs often lack explicit grounding in legal logic and evidence. 
  We present Code-as-Auditor, an LLM-based framework that extends the model’s reasoning capability toward structured and evidence-grounded compliance assessment. 
  The framework translates regulatory information into (1) formalized checklists and executable decision trees, encoding regulations and conditions as interpretable code structures. 
  During inference, each checklist item is (2) dynamically expanded into factual and counterfactual questions, guiding the model to reason over case-specific evidence and potential violations. 
  This process establishes a reasoning pipeline that proceeds from evidence identification, through rule application, to final decision-making, while a self-verification loop improves the logical consistency of the generated code and the traceability of outcomes.
  Experiments on privacy and data protection scenarios demonstrate that Code-as-Auditor delivers more accurate and evidence-backed evaluations, enabling automated compliance regulation checking grounded in explicit regulatory criteria.
\end{abstract}



\maketitle

\section{Introduction}
\begin{figure}[!t]
    \centering
     \includegraphics[width=\columnwidth]{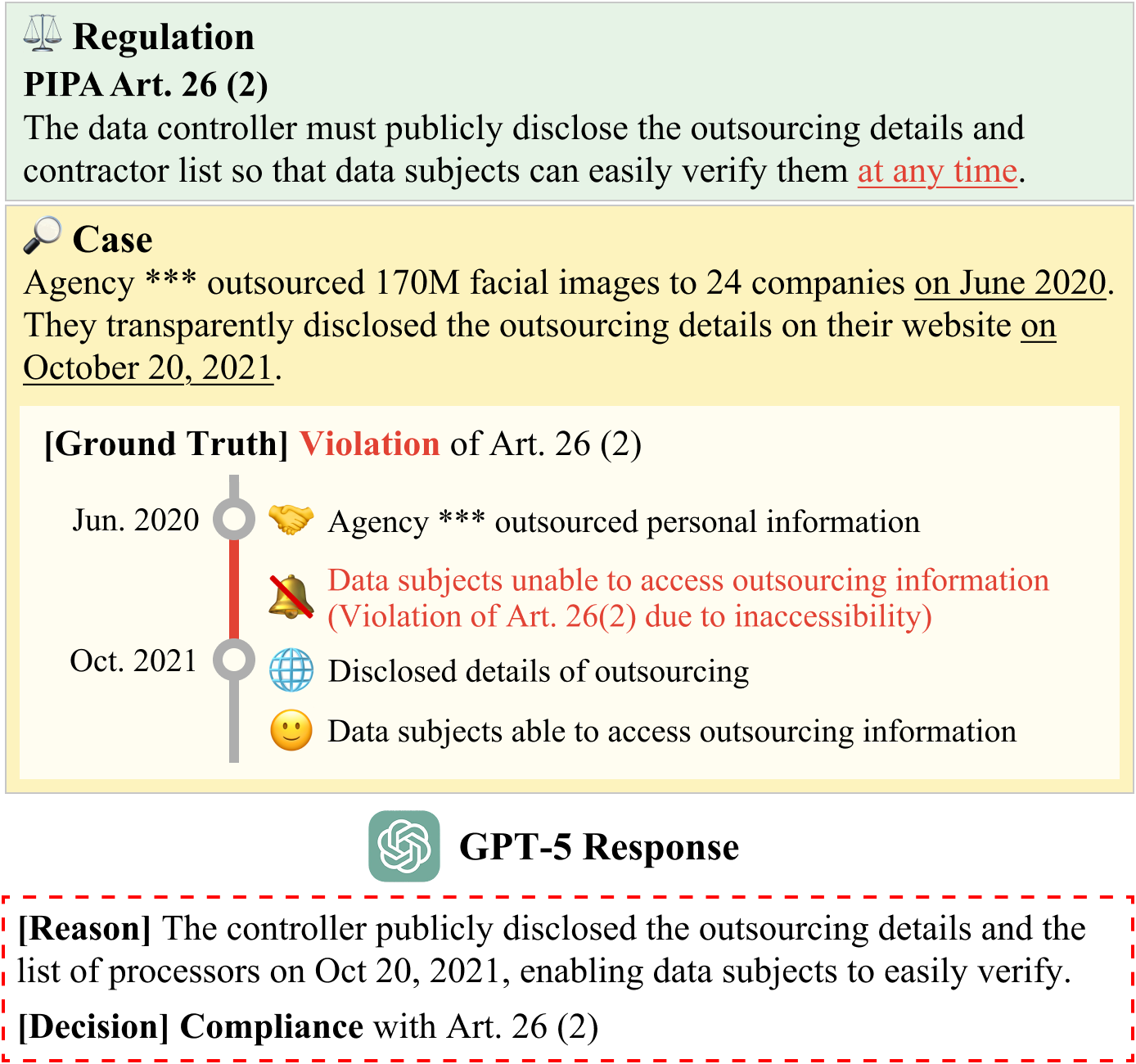}
    \caption{An example of false compliance determined by an LLM, where legality is concluded from disclosure alone, missing the regulation’s temporal conditions.}
    \Description{A compliance example where an LLM concludes that disclosure is sufficient, while the actual regulation also requires temporal conditions to be satisfied.}
    \label{fig:limitation}
\end{figure}

\begin{figure*}[t]
    \centering
    \includegraphics[width=1\textwidth]{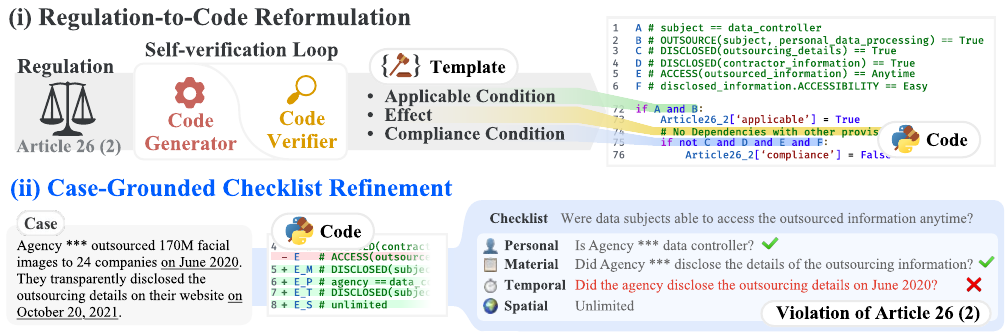}
    \caption{Overview of Code-as-Auditor with empirical results, showing successful detection of the temporal violation missed by LLM in Figure \ref{fig:limitation}.}
    \Description{An overview of the Code-as-Auditor workflow from regulation-to-code reformulation through checklist refinement and compliance reasoning, with example empirical results.}
    \label{fig:overall_framework}
\end{figure*}

Large language model (LLM)-powered regulatory compliance systems have rapidly emerged in recent years, reflecting a broader shift toward automated support for interpreting, organizing, and applying legal and regulatory knowledge~\cite{RAG-based, compass}.
This moved compliance analysis beyond static rule lookup, enabling models to assist with complex documents, evolving requirements, and case-specific judgments at a scale difficult to achieve through manual review alone.
This trend has expanded into contextual privacy analysis, where LLMs are incorporated into systems for assessing privacy-sensitive actions and information flows under specific social and regulatory contexts~\cite{context-reason, privacy-checklist}.
It has also extended to broader legal-assistance systems, where LLMs support practical workflows for consultation, legal-domain reasoning, and data-compliance analysis~\cite{legilm, lawllm}.

Despite this progress, approaches based primarily on open-ended natural-language reasoning remain difficult to rely on for complex regulatory compliance.
Prior analyses distinguish several limitations: legal knowledge may be inaccurate or fabricated~\cite{dahl24}; intermediate reasoning may be unsound even when presented step by step~\cite{Step-by-Step}; and final explanations may fail to reflect expert legal reasoning or provide faithful grounding for the decision~\cite{unpacking, gui2025evaluating}.
These uncertainties arise from implicit rule selection, factual grounding, and inferential steps in natural language form.
As shown in Figure~\ref{fig:limitation}, such systems can therefore produce seemingly plausible yet possibly incorrect legal outputs, while leaving the basis of the compliance decision unverifiable.

%
%
%

Recent neuro-symbolic approaches combine LLMs with explicit symbolic representations and reasoning procedures. The LLM induces candidate rules, which are then encoded in symbolic form and executed by a reasoning engine supporting verifiable decision-making~\citep{privacy-checklist, privaci-bench, RAG-based, tax}.
These approaches have demonstrated the ability of structured compliance reasoning with minimal expert effort. 

However, several limitations remain.
First, despite continued improvements in LLM capability, translating natural language provisions into symbolic rules is still error-prone, frequently producing inconsistencies and logical mistakes~\cite{liu2024exploring}. 
Second, most symbolic representations used in practice lack the expressiveness needed to faithfully model complex regulatory logic, including cross-provision dependencies and extensive exception structures~\citep{goldcoin, horner2025legal, kant2025towards}. 
Lastly, a longstanding semantic gap problem persists between abstract symbolic representations and the ambiguous realities of practical contexts~\cite{ren2024explicit, stengel2023zero}. 

We present \textbf{Code-as-Auditor}, a compliance reasoning framework that converts regulatory provisions into executable code, building on LLMs' capabilities in code reasoning, generation, and evaluation \cite{chain, selfrefine, alphaevolve, llm-as-a-judge}.
The framework formulates regulatory reasoning as LLM-based code generation, translating regulations into a novel, executable code representation, and applying an iterative self-verification loop to improve both logical and syntactic completeness.
The generated code also includes a case-agnostic checklist--a generic set of questions that concretizes the reasoning procedure.
Then, given a specific case, the framework instantiates the checklist by expanding each question through Hans Kelsen’s theory of normative validity \cite{kelsen1967pure}.
This refinement enables compliance reasoning via case-adaptive symbolic execution, systematically aligning abstract regulatory conditions with concrete case facts.

We evaluate Code-as-Auditor across three regulatory frameworks: the European Union Artificial Intelligence Act~\cite{euaiact2024}, the General Data Protection Regulation~\cite{gdpr2016}, and the Personal Information Protection Act (PIPA) of the Republic of Korea~\cite{pipa2011}.
Code-as-Auditor achieves the highest F1-scores across all frameworks, outperforming competing approaches by 3.1 to 12.3 points, demonstrating the effectiveness of executable reasoning for complex compliance assessment.
We additionally evaluate Code-as-Auditor against state-of-the-art reasoning models and generalizable neuro-symbolic baselines in two additional regulatory contexts, the FTC Telemarketing Sales Rule~\cite{telemarketing_sales_rule} and U.S. Federal Tax Law~\cite{us_federal_tax_law}, where it achieves 1.3 and 1.4 accuracy-point gains.
Overall, Code-as-Auditor empirically demonstrates the effectiveness of structured, executable regulatory reasoning for complex compliance assessment.

Our main contributions are as follows:
\begin{enumerate}[label=\textbf{(\arabic*)}]
    \item A novel code-based formalization of regulatory compliance reasoning that operationalizes legal conditions and obligations as executable decision logic.
    \item A verifier-guided regulation-to-code reformulation that improves the structural completeness and logical consistency of the generated decision procedures.
    \item A normatively principled grounding method that aligns abstract regulatory predicates with case-specific facts through
  Kelsen’s normative validity.
    \item A real-world PIPA dataset for evaluating structurally and logically complex compliance reasoning.
\end{enumerate}

\section{Related Work}

\paragraph{Compliance Reasoning}
Prior work on compliance reasoning has explored a range of training-based methods, each learning different aspects of the legal reasoning process. One approach uses \emph{synthetic fine-tuning} on legal instruction data, either constructed via legal syllogism prompting~\cite{disc_lawllm23}, or augmented with synthetic scenarios grounded in contextual integrity theory for privacy statutes~\cite{goldcoin}. Others teach \emph{procedural reasoning patterns} directly, learning explicit syllogistic structure under reinforcement learning~\cite{syler25}. A complementary line trains \emph{input-shaping modules}, such as a deficiency detector that solicits clarifying questions before reasoning, to mitigate underspecified user queries~\cite{ila25}.
However, empirical studies highlight persistent shortcomings of such language-mediated approaches in regulatory tasks. Profiling of LLM legal outputs reports hallucination rates that rise sharply with case complexity, with models often presenting incorrect outputs as confidently as correct ones~\cite{dahl24}. Even with chain-of-thought prompting, models retain alignment gaps between surface fluency and expert legal reasoning, leaving misclassifications uncorrected~\cite{unpacking}, and step-by-step error analyses reveal frequent failures in the soundness and correctness of intermediate reasoning~\cite{Step-by-Step}.

Meanwhile, neuro-symbolic approaches integrate structured symbolic representations to organize regulatory requirements, enabling compliance reasoning without expert intervention. \emph{Contextual integrity theory} is used to represent privacy norms as a checklist of information-flow attributes for violation detection~\cite{privacy-checklist}, followed by a benchmark that evaluates LLM compliance against the same parameters~\cite{privaci-bench}. Other work structures regulatory knowledge into an \emph{eventic graph} and combines it with retrieval-augmented generation to verify business processes against regulations~\cite{RAG-based}. A complementary direction encodes regulatory logic as formal representations: tax statutes are translated into the \emph{Catala} domain-specific language for executable verification~\cite{tax}, and privacy norms are converted into first-order logic predicates through a three-stage pipeline of semantic-role extraction, hierarchical graph construction, and formula synthesis~\cite{quagmire}. However, such formalization remains limited in capturing the structural and logical complexity of legal texts and is prone to hallucinations~\cite{horner2025legal, kant2025towards}.

\paragraph{Code-Based Reasoning}
In contrast, recent progress in LLM coding capability builds on the scaling of code-specific training: pretraining on large multilingual code corpora~\cite{codellama23, deepseekcoder24}, and instruction tuning over evolved code-instruction data~\cite{wizardcoder23}. Beyond data, scaling has extended to post-training, where reinforcement learning with execution-derived rewards refines coding capability~\cite{coderl, deepseek-r1}. Benchmarks have grown correspondingly, moving toward more complex tasks drawn from real-world software engineering~\cite{mbpp, classeval24, jimenez2024swebench}.

Enabled by these advances, recent studies demonstrate that reasoning through executable code improves logical consistency by externalizing intermediate reasoning steps. One approach uses Python-based reasoning---delegating computational steps to an interpreter---and achieves substantial gains on mathematical~\cite{pal} and numerical or financial QA~\cite{pot} benchmarks. Another approach extends code execution with LM-simulated steps, supporting logical, arithmetic, and semantic reasoning tasks~\cite{chain}. A complementary line interleaves reasoning with environmental actions to ground decisions in observable feedback, reducing hallucinations in knowledge-intensive QA and improving success rates in interactive decision-making~\cite{react}. We build on this idea by formulating compliance reasoning as executable code generation.


\paragraph{Structured Normative Representations}
Our framework draws on two complementary theoretical traditions. 
The IRAC framework~\cite{Trautman2025Template} provides a principled sequential structure for legal reasoning, decomposing legal analysis into four ordered stages: \textbf{(1)} the \emph{Issue} articulates the legal question at stake in the case; \textbf{(2)} the \emph{Rule} specifies the law---statutes, doctrines, or precedents---applicable to that issue; \textbf{(3)} the \emph{Application} applies the rule to the specific facts of the case; and \textbf{(4)} the \emph{Conclusion} states the legal outcome that follows from the application. This form of structure is widely employed in computational compliance systems ranging from business process compliance~\cite{Hashmi2016NormativeRequirements} and semantic web-based checking~\cite{Francesconi2022Patterns} to recent LLM-based legal compliance~\cite{Guha2023LegalBench, Jiang2023LegalSyllogism}. We adopt this structure as the basis for our code template.

While IRAC captures the deductive structure of compliance reasoning, bridging abstract regulatory conditions and concrete case facts remains a challenge. Kelsen’s theory of norms \cite{kelsen1967pure} provides a principled framework for this bridging, positing that a norm’s validity is fully characterized only when four dimensions are specified: the \textbf{Personal Sphere} (addressee), \textbf{Material Sphere} (action), \textbf{Temporal Sphere} (timing), and \textbf{Spatial Sphere} (territory). Formally, a norm $p$ decomposes into four sphere boundaries
\begin{equation}
p \;=\; \big(V^{P}_{p},\, V^{M}_{p},\, V^{T}_{p},\, V^{S}_{p}\big),
\end{equation}
and is valid for a case $\omega$ iff every sphere is either unlimited or contains $\omega$:
\begin{equation}
\mathrm{Valid}(p, \omega) \;\Longleftrightarrow\; \bigwedge_{d \in \{P,\,M,\,T,\,S\}} \big(V^{d}_{p} = \mathrm{unlimited} \;\vee\; \omega \in V^{d}_{p}\big).
\end{equation}
Our framework leverages these four spheres as a theoretically grounded ontology, decomposing abstract norms into structured boundaries against which the entailment of a norm by a concrete case is determined.

\begin{figure}[t]
    \centering
     \includegraphics[width=\columnwidth]{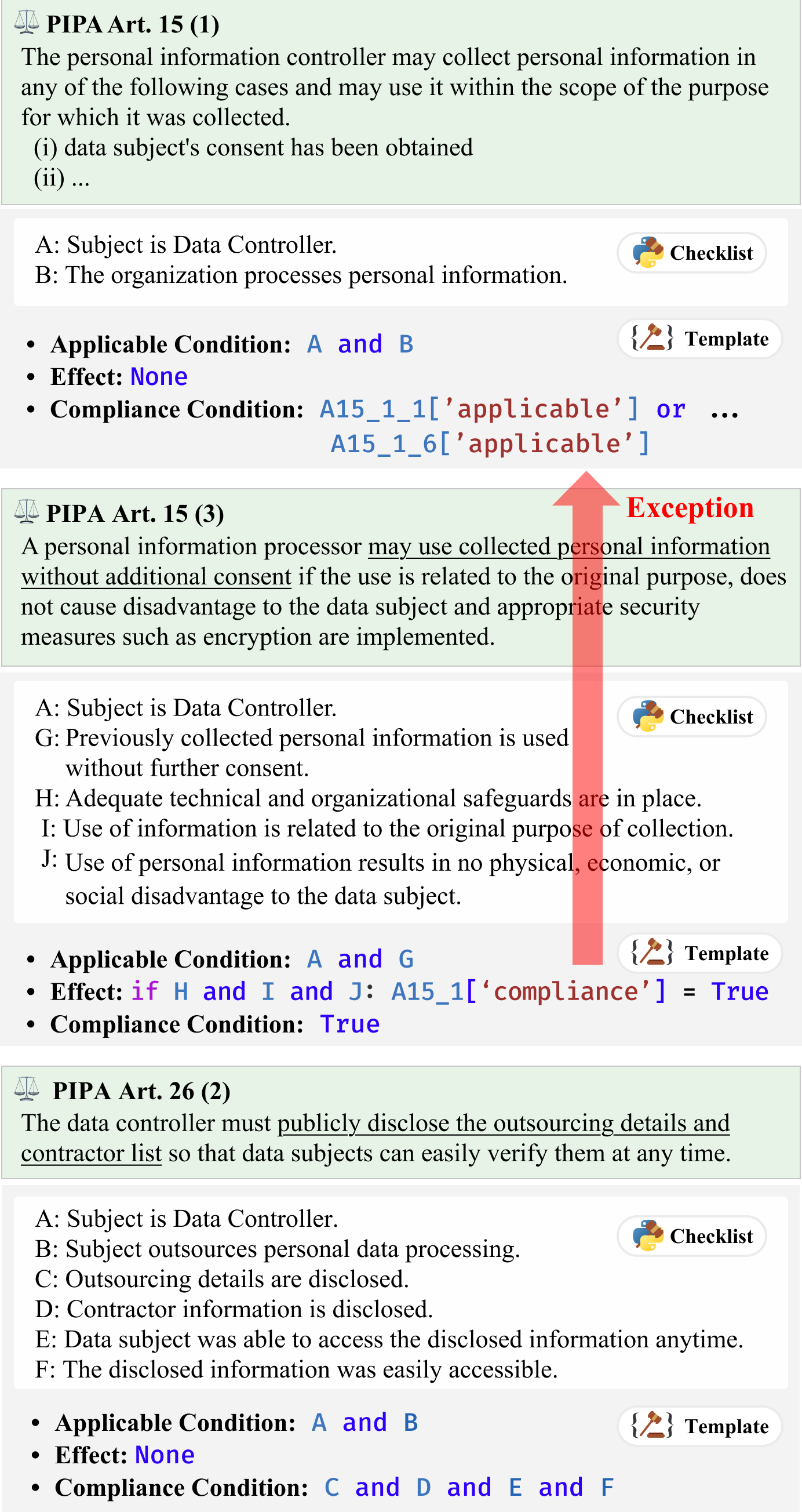}
    \caption{Illustrative examples of the template based on Articles 15(1), 15(3) and 26(2) of the Personal Information Protection Act. Article 15(3) represents an exception to the consent requirement in Article 15(1).}
    \Description{Three side-by-side code blocks showing the executable code template instantiated for PIPA Articles 15(1), 15(3), and 26(2). Each block contains the applicable condition, compliance condition, and effect fields together with the checklist variables they reference, illustrating how a parent provision, its exception, and a cross-referenced provision are encoded as interpretable code structures.}
    \label{fig:example}
\end{figure}

\section{Code-as-Auditor}

We present Code-as-Auditor, a compliance reasoning framework that reconceptualizes regulatory assessment as the execution of explicit regulatory logic over grounded case evidence. 
The key idea is to direct recent advances in LLMs' code generation and reasoning capabilities toward automated legal interpretation, shifting compliance assessment from implicit natural-language inference to an executable process in which legal conditions, exceptions, and effects are represented as code. This design separates two central requirements of regulatory assessment: 
faithfully modeling the abstract structure of regulation and grounding that structure in concrete case facts. 
As illustrated in Figure~\ref{fig:overall_framework}, Code-as-Auditor realizes this design through (i) regulation-to-code reformulation and (ii) case-grounded checklist refinement.

\textbf{Regulation-to-code reformulation} turns regulatory interpretation into an executable reasoning artifact. 
It translates regulatory provisions into a novel code template that separates applicability, compliance, and legal effect, thereby making the structure of legal reasoning explicit prior to case-specific evaluation. 
Since such formalization must preserve conditions, exceptions, and dependencies in complex regulations, Code-as-Auditor builds on the self-verifying generation paradigm~\cite{selfrefine,llm-as-a-judge}, through a loop between the \textit{code generator} ($f_\mathrm{gen}$) and the \textit{code verifier} ($f_\mathrm{vrf}$). 
The resulting template embeds the IRAC reasoning sequence into executable code and is later instantiated for individual cases through the \textit{code executor} ($f_\mathrm{exe}$).

\textbf{Case-grounded checklist refinement} provides the grounding interface between executable regulatory logic and concrete case evidence. 
While regulation-to-code reformulation captures the abstract structure of a provision, compliance assessment depends on whether the facts of a given case entail the regulatory conditions encoded in the executable code. 
The \textit{checklist refiner} ($f_\mathrm{ref}$) therefore expands checklist variables into case-specific questions structured by Kelsen's four spheres of validity~\cite{kelsen1967pure}, aligning each condition with the relevant personal, material, temporal, and spatial dimensions of the case.
This normatively structured grounding reduces the semantic gap between abstract regulatory predicates and concrete factual contexts, enabling Code-as-Auditor to evaluate regulatory entailment over case evidence.

\subsection{Regulation-to-Code Reformulation}\label{subsec:pipeline}

\begin{algorithm}[t]
\caption{Self-verification Loop}
\label{alg:self_verification_loop}
\begin{algorithmic}[1]
\Require Regulation $R$
\State Initialize checklist $\Gamma \gets \emptyset$
\State Initialize code set $\Pi \gets \emptyset$
\For{$p \in R$}
    \State Initialize candidate set $\mathcal{C} \gets \emptyset$
    \State Initialize code and feedback $\pi^{(p)}_0 \gets \emptyset,\ \eta_0 \gets \emptyset$
    \For{$i = 1$ to $N$}
        \State $(\pi^{(p)}_i, \Delta\Gamma_i) \gets f_{\mathrm{gen}}(p, \pi_{i-1}^{(p)}, \eta_{i-1})$
        \State $\eta_i \gets f_{\mathrm{vrf}}(p, \pi^{(p)}_i, \Delta\Gamma_i)$
        \State $\mathcal{C} \gets \mathcal{C} \cup \{(\pi^{(p)}_i, \eta_i)\}$
    \EndFor
    \State $\pi^{*(p)} \gets \underset{(\pi^{(p)}_i, \eta_i) \in \mathcal{C}}{\operatorname{arg\,max}}\ \textsc{score}(\eta_i)$
    \State $\Gamma \gets \Gamma \cup \Delta\Gamma$
    \State $\Pi \gets \Pi \cup \{\pi^{(p)*}\}$
\EndFor
\State \Return $\Pi, \Gamma$
\end{algorithmic}
\end{algorithm}

\begin{algorithm}[t]
\caption{Code Executor $\Psi_{\mathrm{exe}}(\pi^{(p)}, \omega)$}
\label{alg:executor}
\begin{algorithmic}[1]
\Require $\pi^{(p)} = (\tau^{(p)}, \Gamma, \{\pi^{(q)}\}_{q \in \text{child}(p)})$, case $\omega$
\State $r^{(p)} \gets (\neg\alpha, \phi)$ \Comment{initialize return value}
\If{$C_{\text{app}}$}
    \State $r^{(p)} \gets (\alpha, \phi)$ \Comment{provision is applicable}
    \State Execute $\varepsilon$
    \For{each $q \in \text{child}(p)$}
        \State $\Psi_{\mathrm{exe}}(\pi^{(q)}, \omega)$ \Comment{execute child provision}
    \EndFor
    \If{$\neg C_{\text{com}}$}
        \State $r^{(p)} \gets (\alpha, \neg\phi)$ \Comment{provision not complied}
    \EndIf
\EndIf
\State \Return $r^{(p)}$
\end{algorithmic}
\end{algorithm}

\paragraph{Code template.} We introduce the \textit{code template} $\tau$ that encodes a principled compliance reasoning path of IRAC~\cite{Trautman2025Template} directly into code structure:
\begin{equation}
\tau = (C_\mathrm{app}, C_\mathrm{com}, \varepsilon)
\end{equation}
\label{eq:schema}
where $C_\mathrm{app}$ is the \textit{applicable condition}, $C_\mathrm{com}$ is the \textit{compliance condition}, and $\varepsilon$ is the \textit{effect}.
The \textit{applicable condition} and the \textit{compliance condition} are boolean statements that determine whether the regulation applies to a given case and whether the case complies with the regulation, respectively. The effect captures executable interactions with other provisions, enabling the representation of complex and multi-dimensional regulatory outcomes. An illustrative example of the resulting template is provided in Figure~\ref{fig:example}.

\paragraph{Checklist.} Our framework introduces checklists as the mechanism for grounding symbolic regulatory components in case-specific facts. Each checklist item $\gamma \in \Gamma$ serves as an interface that binds abstract predicates in $C_\mathrm{app}$, $C_\mathrm{com}$, and $\varepsilon$ to concrete evidence at test time. Since these variables are defined at the predicate level, the same $\gamma$ may be reused across provisions sharing common conditions, thereby capturing inter-provision dependencies through shared bindings. The cumulative set $\Gamma$ thus encodes both fact-to-rule and rule-to-rule mappings within the regulatory structure. The checklist is accumulated through verifier-guided optimization, with the generator proposing candidate variables and the verifier selecting those that best preserve regulatory logic while supporting case-level grounding.

\begin{figure}[!t]
    \centering
    \includegraphics[width=0.95\columnwidth]{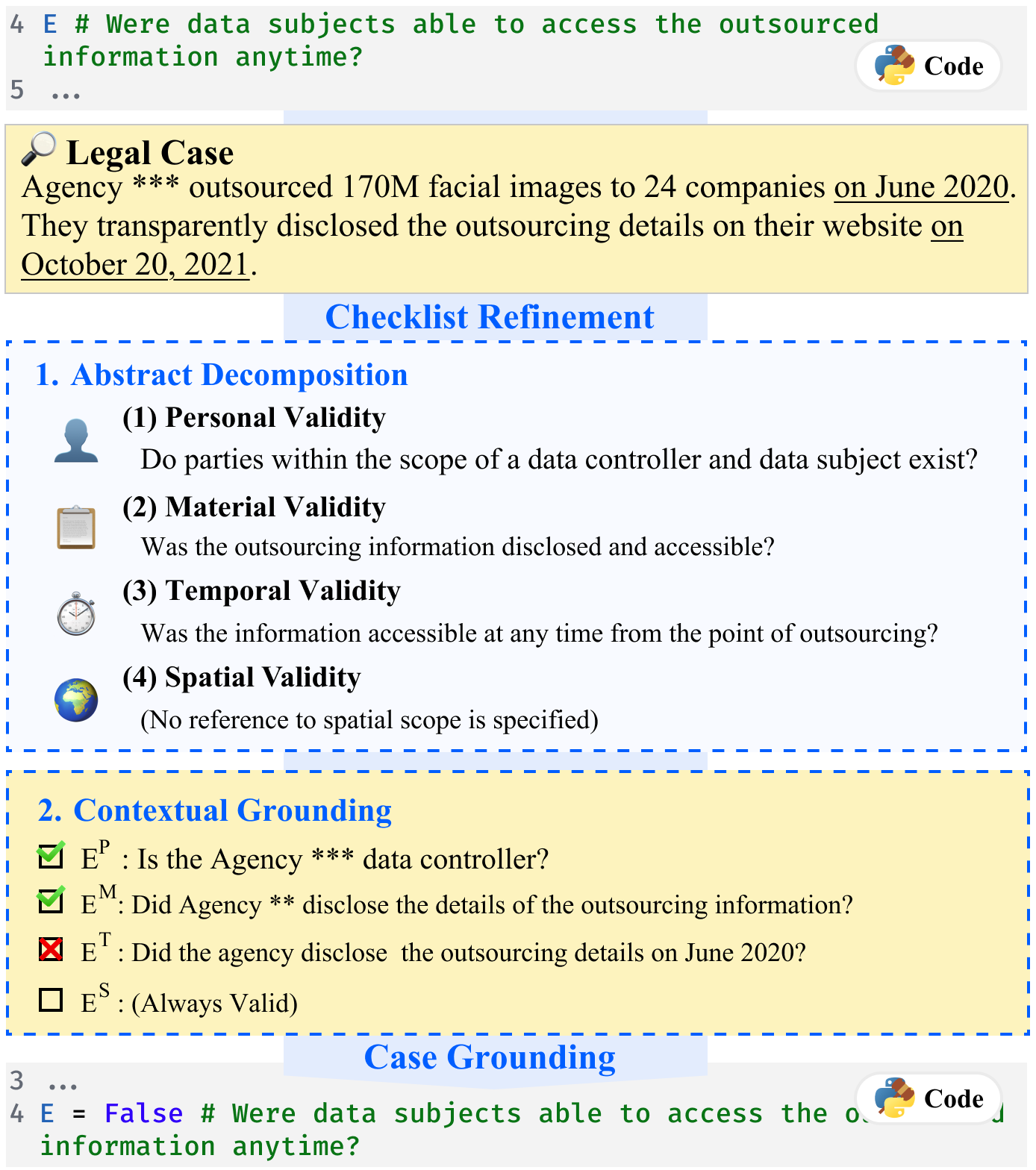}
    \caption{Case-grounded checklist reformulation based on Kelsen's four spheres of validity.}
    \Description{A two-stage diagram showing how an abstract checklist item is refined for a concrete case. The first stage, abstract decomposition, splits the item into four sub-conditions corresponding to Kelsen's personal, material, temporal, and spatial spheres of validity. The second stage, contextual grounding, rewrites each sub-condition into a case-specific question that evaluates to True, False, or None against the given case evidence.}
\label{fig:checklist_reformulation}
\end{figure}

\paragraph{Executable code.} Together with the \textit{code template} $\tau$ and the executable code of its child provisions, the executable code $\pi^{(p)}$ for a provision $p$ is defined as
\begin{equation}
    \pi^{(p)} = (\tau^{(p)},\ \Gamma,\ \{\pi^{(q)}\}_{q \in \operatorname{child}(p)})
\end{equation}
where $\operatorname{child}(p)$ indicates sub-provisions of $p$. This recursive definition reflects the hierarchical structure of regulation. 

Let $R$ denote a regulation consisting of a hierarchy of provisions. The complete executable code $\pi^{(R)}$ is then obtained by composing the executable codes of all provisions, forming a tree structure rooted at the top-level provisions.

\paragraph{Self-verification loop.}
\label{method:verification_loop}
We propose the self-verification loop for structured code generation. This loop iteratively refines the generated code through the \textit{code generator} ($f_\mathrm{gen}$) and the \textit{code verifier} ($f_\mathrm{vrf}$). Let $\Pi^{(p)}$ denote the set of candidate codes generated over $N$ iterations for provision $p$. 
The optimal code is selected as:
\begin{equation}
    \pi^{*(p)} = \operatorname*{arg\,max}_{\pi \in \Pi^{(p)}} \operatorname{score}(f_\mathrm{vrf}(p, \pi))
\end{equation}
where $\operatorname{score}(\cdot)$ quantifies the outcome of self-verification, serving as the objective for deriving the most complete code through repeated verification.

The \textit{code generator} ($f_\mathrm{gen}$) takes a regulatory provision $p$, the prior code $\pi_{i-1}^{(p)}$, 
and previous feedback $\eta_{i-1}$ given from the \textit{code verifier} described below, producing both a refined code and newly proposed checklist variables:
\begin{equation}
    f_\mathrm{gen} : (p, \pi_{i-1}^{(p)}, \eta_{i-1}) \mapsto (\pi_i^{(p)}, \Delta\Gamma_i).
\end{equation}
The \textit{code verifier} ($f_\mathrm{vrf}$) evaluates how accurately the code and proposed checklist represent the regulatory logic returning a feedback $\eta_i$ including both a numerical score and an explanation:
\begin{equation}
    f_\mathrm{vrf} : (p, \pi_i^{(p)}, \Delta\Gamma_i) \mapsto \eta_i.
\end{equation}
The iteration starts from $\pi_0^{(p)} = \emptyset$ and $\eta_0 = \emptyset$. 
The verifier scores each candidate code along five criteria:
\begin{itemize}
    \item \textbf{Necessity}: whether each newly introduced checklist variable is essential or can be replaced by existing variables.
    \item \textbf{Atomicity}: whether each condition is concrete and binary-answerable, avoiding vague vernacular.
    \item \textbf{Hierarchical Integrity}: whether the code respects hierarchical structure of regulation.
    \item \textbf{Logical Completeness}: whether the code faithfully represents regulatory logic and handles None (unspecified) values.
    \item \textbf{Syntactic Validity}: whether the code is syntactically valid and references only defined variables.
\end{itemize}
Each criterion is scored on a 0-5 scale, and the loop terminates when all criteria reach the maximum score or further iterations no longer improve the scores, up to a maximum of $N$ iterations. By default, we set 10 verification loops following the empirical findings in Section~\ref{sec:scale}. Algorithm~\ref{alg:self_verification_loop} summarizes the full procedure, including the accumulation of checklist variables.

The resulting code-form symbolic representation of the regulation serves as an externalized deductive reasoning path that, once generated, is repeatedly reused across cases for evaluation.

\paragraph{Code executor and Compliance decision tree.}

To evaluate compliance systematically, we organize the executable codes $\pi^{(R)}$ into a hierarchical structure termed the \textit{compliance decision tree}, where the \textit{code executor} ($f_\mathrm{exe}$) traverses recursively and executes each provision code $\pi^{(p)}$, realizing compliance assessment as a tree-structured decision process.
For each provision code $\pi^{(p)}$, the executor produces a compliance evaluation result $r^{(p)}$ of the case $\omega$:
\begin{equation}
    f_\mathrm{exe} : (\pi^{(p)}, \omega) \mapsto r^{(p)}, \quad r^{(p)} \in \mathcal{A} \times \mathcal{P}
\end{equation}
where $\mathcal A = \{\alpha, \neg\alpha\}$ and $\mathcal P = \{\phi, \neg\phi\}$, $\alpha$ denotes applicability and $\phi$ denotes compliance.
The executor recursively transitions each node to $\alpha$ if its \textit{applicable condition} is satisfied, and subsequently to $\neg\phi$ if its \textit{compliance condition} is not met. The full execution procedure is described in Algorithm~\ref{alg:executor}.

After executing the complete executable code $\pi^{(R)}$, any provision with an execution result of $(\alpha, \neg\phi)$—indicating that the provision is applicable but not complied with—is judged as violated, where $\operatorname{Violation}(p, \omega)$ denotes whether provision $p$ is determined to be violated for the case $\omega$ by code execution:
\begin{equation}
\operatorname{Violation}(p, \omega)
\Leftrightarrow
f_\mathrm{exe}(\pi^{(p)}, \omega)\!=\!(\alpha, \neg\phi).
\end{equation}
Violations are then propagated upward along the provision hierarchy to the regulation $R$, which is considered compliant if and only if no violated provision exists.

\subsection{Case-Grounded Checklist Refinement}\label{subsec:checklist_reform}

The \textit{checklist refiner} ($f_\mathrm{ref}$) rewrites the original checklist $\Gamma$ based on the case $\omega$ to produce a case-grounded checklist $\Gamma'$:
\begin{equation}
    f_\mathrm{ref} : (\Gamma, \omega) \mapsto \Gamma'.
\end{equation}

This refinement proceeds in two stages, as illustrated in Figure~\ref{fig:checklist_reformulation}: \textit{abstract decomposition}, which breaks each checklist item into normatively grounded sub-conditions, followed by \textit{contextual grounding}, which grounds each sub-condition to the specific entities and context of the case.

In the \textit{abstract decomposition} stage, each checklist item $\gamma$ is decomposed into four sub-conditions according to Kelsen's four spheres of validity: the responsible party (personal), required action (material), time constraint (temporal), and applicable jurisdiction (spatial).

In the \textit{contextual grounding} stage, each decomposed sub-condition is rewritten regarding the case $\omega$, yielding case-specific questions for each of the four normative dimensions. To handle evidentiary uncertainty, we adopt a ternary evaluation logic, where each sub-question evaluates to \texttt{True} (support), \texttt{False} (contradict), and \texttt{None} (absent).
This design reflects record-bound adjudication: absent facts are treated as unavailable evidence rather than contradiction.    
In application of automated regulatory auditing systems, \texttt{None} can function as a signal of unresolved information required for compliance judgment.

\begin{table}[t]
\centering
\caption{Abbreviated prompts for ($f_\mathrm{gen}$) and ($f_\mathrm{vrf}$).
}
\begin{tabular}{@{}p{0.95\columnwidth}@{}}
\toprule
\multicolumn{1}{@{}l@{}}{\textbf{Code Generator} \; ($f_\mathrm{gen}$)} \\
\midrule
You encode each article of \{LAW\} as a JSON schema that detects whether the subject complies.\\[1pt]
\textit{Objectives.}
\begin{itemize}
\item Generate a logically consistent JSON schema for the target legal unit.
\item Prefer existing variables and add only necessary variables for evaluation.
\end{itemize} \\
\textit{Rules.}
\begin{itemize}
\item Each unit is encoded with \texttt{applicable\_condition}, \texttt{effect\_code}, and \texttt{compliance\_condition}.
\item If \texttt{applicable\_condition} is \texttt{False}, the unit is skipped.
\item Use \texttt{effect} for cross-article side effects or exceptions.
\item Added variables must be atomic, observable, and handle \texttt{None} explicitly.
\end{itemize} \\

\midrule
\multicolumn{1}{@{}l@{}}{\textbf{Code Verifier} \; ($f_\mathrm{vrf}$)} \\
\midrule
You score a generated JSON encoding of one article of \{LAW\}. \\[1pt]
\textit{Criteria} (each scored $0$--$5$).
\begin{itemize}
\item \textbf{Necessity}-added variables are needed and not duplicates.
\item \textbf{Specificity}-descriptions are concrete, declarative, and Yes/No-answerable.
\item \textbf{Logic}-conditions reflect the article, exceptions, and \texttt{None} handling.
\item \textbf{Code}-Python is syntactically valid and references only defined names.
\item \textbf{Hierarchy}-higher-level units do not depend on lower-level states.
\item For scores below $5$, identify specific issues and one-line fixes.
\end{itemize} \\[1pt]
\bottomrule
\end{tabular}
\label{tab:prompts}
\end{table}

\begin{table}[t]
\centering
\caption{Article counts and internal cross-references per regulation. PIPA shows the highest cross-reference density.}
\begin{tabular}{l|ccc}
\toprule
& \textbf{EU AI Act} & \textbf{GDPR} & \textbf{PIPA} \\
\midrule
Total Articles & 113 & 99 & 76 \\
Cross-References & 69 & 79 & 85 \\
\midrule
\textit{Cross-References / Articles} & 0.61 & 0.80 & 1.12 \\
\bottomrule
\end{tabular}
\label{tab:document-stats}
\end{table}

\begin{table}[t]
\centering
\caption{Deontic statements and conditional branches per regulation, with PIPA exhibiting the highest density.}
\begin{tabular}{l|ccc}
\toprule
& \textbf{EU AI Act} & \textbf{GDPR} & \textbf{PIPA} \\
\midrule
Deontic & 1,683 & 986 & 353 \\
Conditional & 696 & 592 & 952 \\
\midrule
\textit{Conditional / Deontic} & 0.41 & 0.60 & 2.70 \\
\bottomrule
\end{tabular}
\label{tab:deontic-stats}
\end{table}

Through this two-stage process, each $\gamma$ is refined into a conjunction of case-specific inquiries, and a checklist is deemed entailed by the case only when all of its sub-conditions evaluate to \texttt{True} within their respective normative scopes.

\subsection{Implementation Prompts}
\label{subsec:implementation_prompts}

Table~\ref{tab:prompts} specifies the implementation prompts for the \textit{code generator} ($f_\mathrm{gen}$) and \textit{code verifier} ($f_\mathrm{vrf}$), introduced in the regulation-to-code reformulation of Section~\ref{subsec:pipeline}.

\section{PIPA Evaluation Dataset}
\label{sec:pipa_dataset}

\begin{figure}[t]
    \centering
     \includegraphics[width=0.75\columnwidth]{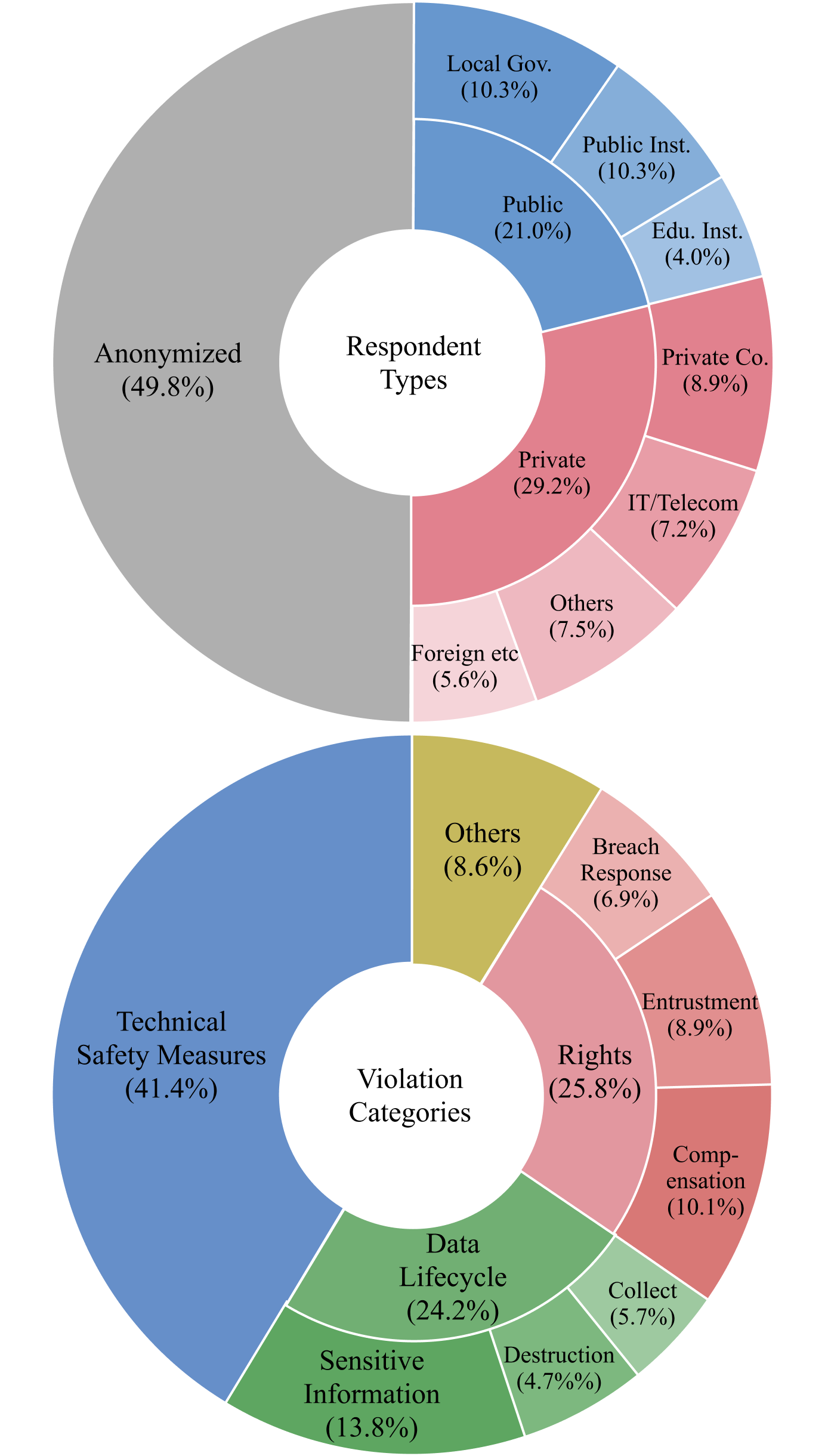}
    \caption{Distribution of the real-world PIPA evaluation dataset, across respondent types (top) and violation categories (bottom).}
    \Description{Two nested donut charts summarizing the PIPA evaluation dataset. The top chart shows respondent types: Anonymized (49.8\%), Public sector (21.0\%, comprising Local Government 10.3\%, Public Institution 10.3\%, and Educational Institution 4.0\%), and Private sector (29.2\%, comprising Private Company 8.9\%, IT/Telecom 7.2\%, Foreign and others 5.6\%, and Others 7.5\%). The bottom chart shows violation categories: Technical Safety Measures (41.4\%), Rights-related provisions (25.8\%, comprising Compensation 10.1\%, Entrustment 8.9\%, and Breach Response 6.9\%), Data Lifecycle (24.2\%, including Collect 5.7\% and Destruction 4.7\%), Sensitive Information (13.8\%), and Others (8.6\%); cases may belong to multiple categories.}
    \label{fig:dataset_statistics}
\end{figure}

As shown in Figure~\ref{fig:dataset_statistics}, we construct an evaluation dataset of real-world compliance cases that requires reasoning over a regulation of substantial structural complexity. We build the dataset on the Personal Information Protection Act (PIPA) of the Republic of Korea, which exhibits comparatively higher structural complexity (Table~\ref{tab:document-stats}) and logical complexity (Table~\ref{tab:deontic-stats}). The dataset comprises 406 publicly available deliberation decisions issued by the Personal Information Protection Commission (PIPC).\footnote{\url{https://www.pipc.go.kr/np/}}

\subsection{Preprocessing and Verification}
Every collected document was first manually inspected to confirm that personally identifiable information had been properly anonymized. We then used GPT-5~\cite{gpt5} to compress each decision into a concise case description containing central legal points at issue in the case.  Compliance labels are taken directly from the Commission's adjudications. To maintain high data quality, authors cross-checked every summary against its source to ensure factual consistency with the original deliberation decision.


\begin{table*}[t]
\centering
\caption{Overall performance comparison across three regulatory frameworks evaluated with the Qwen2.5-7B model.}
\begin{tabular}{lc>{\columncolor{gray!20}}c c>{\columncolor{gray!20}}c c>{\columncolor{gray!20}}c}
\toprule
 & \multicolumn{2}{c}{EU AI Act} & \multicolumn{2}{c}{GDPR} & \multicolumn{2}{c}{PIPA} \\
\cmidrule(lr){2-3} \cmidrule(lr){4-5} \cmidrule(lr){6-7}
Methods          & Recall   & \cellcolor{white}F1-score & Recall    & \cellcolor{white}F1-score & Recall      & \cellcolor{white}F1-score   \\
\midrule
Direct Prompting & 96.0\std{0.2} & 46.0\std{0.1} & 72.8\std{0.2}  & 43.5\std{0.1} & 87.5\std{1.9}  & 29.4\std{0.6}   \\
CI Parameter ~\cite{privacy-checklist} & 87.4\std{0.5} & 77.3\std{0.7} & 55.9\std{0.5}  & 57.8\std{0.3} & 27.6\std{0.0}  & 31.7\std{0.4}   \\
Deontic Triplet ~\cite{RAG-based} & 80.4\std{0.4} & 69.0\std{0.4} & 43.8\std{0.1}  & 43.0\std{0.1} & 32.7\std{1.0}  & 29.4\std{0.9}   \\
PolicyLR ~\cite{policylr} & 51.3\std{0.1} & 48.0\std{0.1} & 48.6\std{10.7} & 47.0\std{5.7} & 60.8\std{5.5}  & 40.1\std{2.2}   \\
Semantic FOL ~\cite{quagmire} & 82.9\std{0.8} & 87.6\std{0.1} & 54.2\std{1.5}  & 51.9\std{1.6} & 60.5\std{1.0}  & 43.0\std{0.7}   \\
Code-as-Auditor (Ours) & 84.5\std{1.4} & \textbf{90.7}\std{0.5} & 74.7\std{4.0}  & \textbf{61.2}\std{0.9} & 80.0\std{5.1}  & \textbf{55.3}\std{3.6}   \\
\bottomrule
\end{tabular}
\label{tab:main_table}
\end{table*}

\section{Experiments}
\subsection{Experimental Settings}
\label{sec:experimental_settings}
\paragraph{Datasets.}
We evaluate on three regulatory frameworks: the EU AI Act~\cite{euaiact2024}, GDPR~\cite{gdpr2016}, and PIPA~\cite{pipa2011}. For the EU AI Act and GDPR, we utilize benchmarks from PrivaCI-Bench~\cite{privaci-bench}, which comprises real court cases, privacy policies, and synthetic vignettes built from official compliance toolkits. 
For PIPA, evaluation is conducted on the dataset we construct in Section~\ref{sec:pipa_dataset}.
We arrange the evaluation across the EU AI Act, GDPR, and PIPA as an ordered spectrum of complexity.

\paragraph{Metrics.}
We evaluate compliance reasoning at the article level across all test cases. We report micro-averaged recall and F1-score, evaluated over three trials, defining a true positive as a correctly identified violation of a specific article within a given case.

\paragraph{Baselines.} 
We compare Code-as-Auditor against five baseline methods that utilize LLM-generated symbolic representations for compliance reasoning: 
(1) \textbf{Direct Prompting}, a basic baseline without intermediate representations; 
(2) \textbf{CI Parameter}~\cite{privacy-checklist}, which models information flows as 5-tuples based on Contextual Integrity \cite{contextual_integrity};
(3) \textbf{Deontic Triplet}~\cite{RAG-based}, which parses regulations into $\langle$\textit{agent, deontic operator, action}$\rangle$ triplets structured via an eventic knowledge graph; 
(4) \textbf{PolicyLR}~\cite{policylr}, which represents policies as valuations over atomic formulae; and 
(5) \textbf{Semantic FOL}~\cite{quagmire} encoding regulations as first-order logic predicates.

\paragraph{Implementation Details.}

We employ GPT-5~\cite{gpt5} to generate the symbolic representations for all baselines and Code-as-Auditor.
For the inference stage, Qwen2.5-7B~\cite{qwen2_5} serves as the default backbone to evaluate compliance across all methods.

\subsection{Overall Performance} 
\label{sec:overall_performance}

As shown in Table~\ref{tab:main_table}, tuple-based methods such as CI Parameter and Deontic Triplet perform well on simpler regulations but suffer substantial degradation as complexity increases (CI Parameter: $-$45.6; Deontic Triplet: $-$39.6), whereas logic-based representations maintain comparatively stronger performance even under higher complexity, achieving 40.1 (PolicyLR) and 43.0 (Semantic FOL) F1-scores on PIPA. Code-as-Auditor achieves the highest F1-scores across all frameworks, surpassing the second-best methods by +3.1 on the EU AI Act, +3.4 on GDPR, and +12.3 on PIPA, demonstrating its well-structured hierarchical design and high expressive capacity consistently leading to high performance under increasing complexity.

\subsection{Analysis}

\subsubsection{Inference Model Scale}
\label{sec:inference_model_scale}
To examine the sensitivity of symbolic representations to inference model capacity, we evaluate tuple-based (CI Parameter, Deontic Triplet) and logic-based methods (PolicyLR, Semantic FOL) using smaller inference models (Qwen2.5-1.5B and 0.5B), with results summarized in Table \ref{tab:param_efficiency}. The effect is particularly pronounced for logic-based representations, where reduced inference capacity widens the semantic gap between formalized rules and case facts, leading to larger performance degradation.
In comparison, Code-as-Auditor externalizes the reasoning path and reduces inference to evaluating only atomic conditions, resulting in greater robustness to model scaling.
Code-as-Auditor preserves the highest F1-score while exhibiting the smallest degradation compared to the 7B model in the EU AI Act ($-$4.7) and PIPA ($-$12.8). 

\begin{table}[t]
\centering
\caption{The overall F1 score across three regulatory frameworks with Qwen2.5-1.5B and 0.5B model.}
\begin{tabular}{llcc}
\toprule
\textbf{Regulation} & \textbf{Method} & \textbf{1.5B} & \textbf{0.5B} \\
\midrule
\multirow{3}{*}{EU AI Act} & Tuple& 63.5\std{0.2}  & 49.4\std{0.6}  \\
 & Logic& 39.1\std{0.1}  & 27.2\std{0.2}  \\
 & Ours & 89.9\std{0.8}  & 86.0\std{0.2}  \\
\midrule
\multirow{3}{*}{GDPR} & Tuple& 32.3\std{1.5}  & 25.9\std{1.0}  \\
 & Logic & 21.6\std{0.6}  & 14.2\std{0.3}  \\
 & Ours& 48.7\std{2.1}  & 47.8\std{1.6}  \\
\midrule
\multirow{3}{*}{PIPA} & Tuple& 20.7\std{0.2}  & 18.3\std{0.1}  \\
 & Logic& 16.0\std{0.5}  & 9.6\std{0.4}  \\
 & Ours& 53.8\std{1.1}  & 42.5\std{1.0}  \\
\bottomrule
\end{tabular}
\label{tab:param_efficiency}
\end{table}

\begin{figure}[t]
    \centering
    \includegraphics[width=\columnwidth]{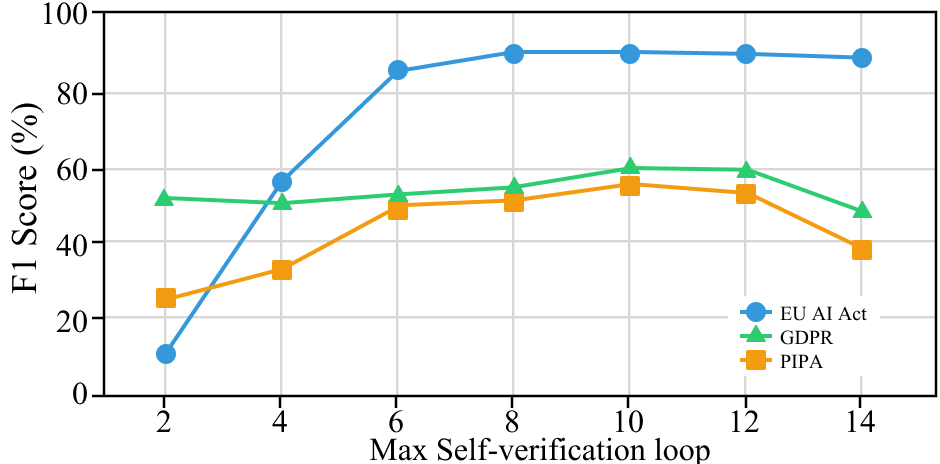}
    \caption{Performance comparison across maximum self-verification loop depth.}
    \Description{Line plots of F1-score as a function of the maximum number of self-verification iterations for the EU AI Act, GDPR, and PIPA. Performance rises with the number of iterations, plateaus around ten iterations, and then declines as additional iterations are added.}
\label{fig:analysis_loop}
\end{figure}

\subsubsection{Self-Verification Loop Scale} \label{sec:scale}
To analyze the effect of the self-verification loop, we experiment by varying the number of verification iterations, with the results shown in Figure~\ref{fig:analysis_loop}. 
Overall performance improves as iteration increases and saturates around 10 iterations, achieving  F1-scores in all frameworks. 

This saturation behavior demonstrates that self-verification effectively guides the model toward logically and syntactically complete code generation. However, additional iterations lead to performance degradation across datasets (AI Act $-$1.5, GDPR $-$12.7, PIPA $-$16.3 at 14 iterations). Consistent with prior work on iterative self-refinement~\cite{chen-etal-2025-magicore}, additional iterations in our legal setting encourage overly deductive compliance reasoning, increasing the semantic gap between symbolic representations and real-world contexts.

\subsubsection{Analysis of Checklist Refinement} \label{sec:checklist_reformulation}
We analyze alternative approaches to ground case evidence on real-world PIPA deliberation cases. We compare Kelsen's four spheres of validity~\cite{kelsen1967pure} against four baselines:
Claim-Element analysis~\cite{medicus2024buergerliches}, a requirement-based legal decomposition, and three factual grounding methods---5W1H, FActScore~\cite{min2023factscore}, and ClaimDecomp~\cite{chen2022claimdecomp}.

\begin{figure}[t]
    \centering
    \includegraphics[width=1\columnwidth]{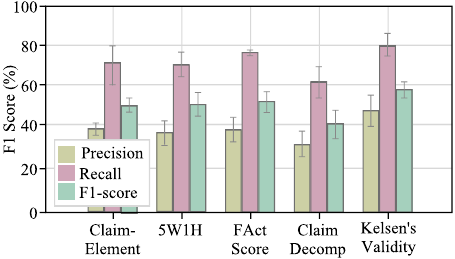}
    \caption{Performance comparison across checklist refinement methods on real-world PIPA cases.}
    \Description{Grouped bar chart comparing five checklist refinement methods on real-world PIPA cases: Claim-Element analysis, 5W1H, FActScore, ClaimDecomp, and Kelsen's four spheres of validity (ours). Each method is reported on precision, recall, and F1, with Kelsen's validity yielding the highest values across all three metrics.}
    \label{fig:symbolic_adaptation}
\end{figure}

Results are illustrated in Figure~\ref{fig:symbolic_adaptation} \textbf{Claim-Element} achieves comparatively high precision through legal requirement-based decomposition (37.44), yet its limited alignment with case evidence constrains overall reasoning performance.
Among fact-oriented methods, \textbf{FActScore} attains high evidence coverage, achieving the second-strongest recall (76.90) and the second-best overall performance (50.21), though its atomic fact decomposition remains weakly aligned with normative rule constraints.
\textbf{Kelsen’s validity (ours)} achieves the best performance across all metrics by decomposing regulatory conditions into normatively principled dimensions, enabling precise rule application while maintaining broad evidence coverage.

\subsubsection{Ablation Study}
\label{sec:ablation_study}
\begin{figure}[t]
    \centering
    \includegraphics[width=1\columnwidth]{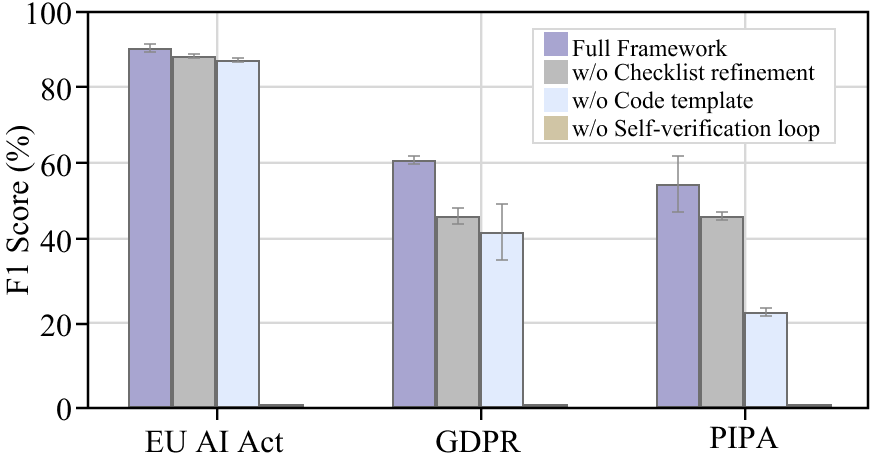}
    \caption{Ablation analysis of Code-as-Auditor. Removing Self-reflection loop resulted in a syntax error.}
    \Description{Bar charts contrasting the full Code-as-Auditor with three ablated variants—without checklist refinement, without the code template, and without the self-verification loop—across the EU AI Act, GDPR, and PIPA. Each bar reports the mean score over three trials with error bars; ablating the self-verification loop drops every score to zero because the generated code is no longer syntactically valid.}
    \label{fig:analysis_ablation}
\end{figure}

We systematically evaluate the contribution of each component by removing it from the full framework. Figure~\ref{fig:analysis_ablation} reports the mean and standard deviation over three randomized trials. 

\paragraph{Checklist refinement.}
Without checklist refinement, recall drops substantially across GDPR ($-$18.65) and PIPA ($-$37.91). This indicates that the symbolic rules fail to bridge the semantic gap between abstract predicates and concrete case facts—a capacity uniquely provided by Kelsen's four spheres of validity.

\paragraph{Code template.}
The impact of removing the code template scales with regulatory complexity: the F1 drop is modest on the EU AI Act ($-$1.57) but severe on PIPA ($-$30.86). This highlights its effectiveness, particularly under complex conditions. 

\paragraph{Self-verification loop.}
Removing the self-verification loop resulted in syntactically invalid code across all three regulatory domains, yielding zero scores on every metric. 
This indicates that the self-verification loop is essential for ensuring executable code generation, serving as a structural prerequisite for the compliance reasoning pipeline.


\subsection{Framework Generality}
\label{sec:generality}
To examine whether Code-as-Auditor generalizes as a methodology beyond privacy-specific compliance, we re-instantiate the full pipeline on two tasks from LegalBench~\cite{Guha2023LegalBench} drawn from regulations disjoint from our primary benchmark: \emph{Telemarketing}, a rule-application task that evaluates violations of the FTC Telemarketing Sales Rule~\cite{telemarketing_sales_rule}, and \emph{SARA Entailment}, a statutory entailment task grounded in U.S. Federal Tax Law~\cite{us_federal_tax_law}. The framework is compared against two baseline families: general-purpose SOTA reasoning models (GPT-5~\cite{gpt5}, Claude 4~\cite{claude}) and generalizable neuro-symbolic methods (Deontic Triplet~\cite{RAG-based}, Semantic FOL~\cite{quagmire}). Results are summarized in Table~\ref{tab:appendix_reasoning}.

Code-as-Auditor’s gains over both backbone models clarify the source of its performance. By outperforming GPT-5, it shows that structured regulatory reasoning improves upon direct reasoning by the artifact-construction model. By outperforming Qwen2.5-7B, it shows that the test-time model benefits from structured neuro-symbolic guidance. Its superiority over other generalizable baselines further demonstrates that the proposed formulation provides a more effective representation for regulatory compliance reasoning. Together, these results demonstrate the broader applicability of Code-as-Auditor to regulatory entailment reasoning across diverse domains and task settings.

\begin{table}[h]
\centering
\caption{Comparison of state-of-the-art reasoning models, generalizable neuro-symbolic approaches, and Code-as-Auditor on the SARA and Telemarketing benchmarks. Averaged accuracy (\%) is reported over three runs.}
\begin{tabular}{l*{2}{>{\centering\arraybackslash}p{2.15cm}}}
\toprule
Reasoning Approach & Telemarketing & SARA \\
\midrule
GPT-5~\cite{gpt5}                 & 90.7\std{1.2} & 78.8\std{0.7} \\
GPT-5 Mini~\cite{gpt5}            & 73.0\std{3.2} & 81.3\std{1.1} \\
Claude Sonnet 4.5~\cite{claude}   & 85.1\std{5.6} & 63.8\std{0.2} \\
Claude Opus 4.5~\cite{claude}     & 84.2\std{5.1} & 57.6\std{2.3} \\
\midrule
Qwen2.5-7B~\cite{qwen2_5}         & 44.6\std{9.1} & 40.3\std{1.6} \\
Deontic Triplet~\cite{RAG-based}  & 56.0\std{3.6} & 54.9\std{1.7} \\
Semantic FOL~\cite{quagmire}      & 59.6\std{0.0} & 56.3\std{1.0} \\
Code-as-Auditor (Ours)            & \textbf{92.0\std{0.4}} & \textbf{82.7\std{3.1}} \\
\bottomrule
\end{tabular}
\label{tab:appendix_reasoning}
\end{table}

\begin{figure*}[t]
    \centering
    \includegraphics[width=0.94\textwidth]{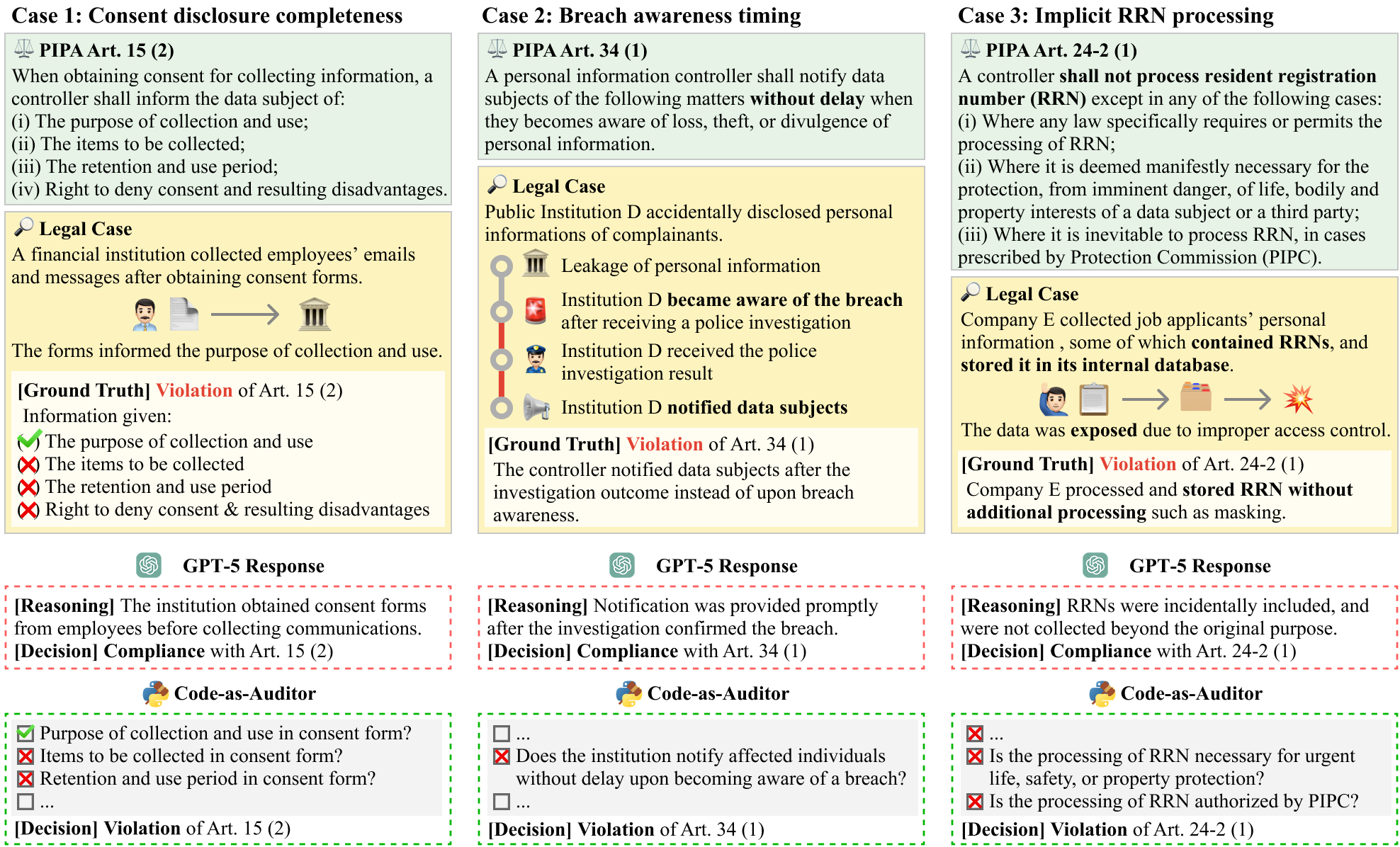}
    \caption{Three challenging compliance reasoning scenarios where GPT-5 misinterprets the legal basis, while Code-as-Auditor correctly identifies the violation.}
    \Description{A three-panel figure presenting qualitative cases side by side. The left panel covers multi-granular logical reasoning, where a consent form satisfied one item of PIPA Article 15(2) but omitted the retention period and the right to refuse. The center panel covers multi-dimensional temporal assessment, where the without-delay notification obligation under Article 34(1) was triggered four months earlier than the actual notice. The right panel covers context-robust evaluation, where a breach-framed case obscured Article 24-2(1) prohibiting unauthorized processing of resident registration numbers. Each panel contrasts GPT-5's misinterpretation with Code-as-Auditor's correct violation detection.}
    \label{fig:analysis_qualitative}
\end{figure*}

\section{Qualitative Evaluation}
\label{sec:qualitative}
Figure~\ref{fig:analysis_qualitative} illustrates three recurring challenges in compliance reasoning, and shows how Code-as-Auditor addresses each.

\paragraph{Multi-granular logical reasoning.}

Provisions with hierarchical structure and enumerated mandatory items are challenging, as they require both holistic and fine-grained evaluation.
A single compliance signal can mask the absence of other required elements.
In Case~1 (left panel), a consent form was obtained from employees but omitted the retention period and the right to refuse, both required under Article 15(2).
Code-as-Auditor decomposed each provision into atomic checklist variables, ensuring that every mandatory item is independently verified.

\paragraph{Multi-dimensional assessment.}
Obligations carry implicit validity dimensions: not only \emph{what} must be done but also \emph{when}, \emph{by whom}, and \emph{where}. 
Reducing compliance reasoning to a single sequential reasoning path may obscure these dimensions by collapsing them.
In Case~2 (center panel), data subjects were notified shortly after investigation results arrived, yet the `without delay' obligation of Article 34(1) had already been triggered four months earlier when the breach was first reported by police.
Code-as-Auditor grounds each checklist item in its relevant validity dimension, successfully binding temporal conditions to the legally defined trigger point.

\paragraph{Context-robust evaluation.}
LLMs frequently exhibit variability depending on prompt formulation and contextual framing \cite{llm_bias_1, llm_bias_2}. When a case has a dominant narrative framing, attention may concentrate on provisions that fit that frame while upstream prohibitions go unexamined.
In Case~3 (right panel), the case is framed as a data breach incident, drawing attention to safety measure violations. Article 24-2(1), which prohibits processing resident registration numbers without a legal exception, falls outside that frame and was not identified.
Code-as-Auditor evaluates each provision by executing its compliance condition code independently. The result is determined by whether the coded conditions are satisfied, not by the narrative context.

\section{Conclusion}

We introduced Code-as-Auditor, which reconceptualizes legal reasoning as code generation and structured reasoning. 
By externalizing reasoning into interpretable code, the approach addresses limitations of language-based compliance and further mitigates several limitations previously observed in neuro-symbolic compliance systems.
Beyond addressing these limitations, structuring compliance reasoning as interpretable and executable code offers a promising pathway toward systematic regulatory auditing.

\section{Discussion}
\paragraph{Analogical Reasoning.}
Real-world legal reasoning rests on both deductive rule application and analogical inference from precedent and expert judgment. This work lays the groundwork for structured reasoning in deductive legal inference, upon which future studies can align reasoning traces with expert preferences~\cite{lahlou} and refine the executable code via precedent-grounded verifiable rewards~\cite{alphaevolve}.

\paragraph{Verifiable Logic.}
Our self-verification loop certifies that a provision is encoded as a well-formed decision procedure, but not that the encoding preserves the meaning of the statutory text. Future work may close this gap by yielding executable regulatory logic that is principled and verifiable~\cite{decomposed, lambada}.

\paragraph{Missing Evidence.}
Following the record-bound maxim \emph{quod non est in actis, non est in mundo}, our ternary logic resolves unproven conditions against violation as the burden of proof dictates, though it does not distinguish genuine record silence from facts that grounding fails to extract. Since each \texttt{None} is bound to an atomic checklist question, the \texttt{None} set enumerates the information still required for a judgment, enabling targeted review in deployment.

\paragraph{Dataset Bias.}
The PIPA dataset is collected from publicly released PIPC deliberation decisions, with compliance labels taken directly from the Commission's adjudications and every summary cross-checked by the authors against its source record. However, as the case descriptions are GPT-5 summaries of already-adjudicated decisions, they may smooth the distractors and buried facts of raw records and underrepresent borderline situations.

\begin{acks}
This work was supported by 
the TIPS(Tech Incubator Program for Startup) R\&D Program (RS-2024-00508880), funded by the Ministry of SMEs, Republic of Korea, 
Institute of Information \& communications Technology Planning \& Evaluation(IITP) grant funded by the Korea government(MSIT) (RS-2019-II190421, AI Graduate School Support Program(Sungkyunkwan University), 
RS-2022-II221045 (2022-0-01045), Self-directed multi-modal Intelligence for solving unknown, open domain problems, 
RS2022-II220043, Adaptive Personality for Intelligent Agents, 
No.RS-2025-25442569, AI Star Fellowship Support Program(Sungkyunkwan Univ.), 
RS-2025-02218768, Accelerated Insight Reasoning via Continual Learning), 
Samsung Electronics Co., Ltd, 
Institute of Information \& Communications Technology Planning \& Evaluation(IITP)-ITRC(Information Technology Research Center) grant funded by the Korea government(MSIT) (IITP-2026-RS-2024-00437633), 
National Research Foundation of Korea (NRF) grant funded by the Korea government (MSIT) (No. RS-2026-25474409).
\end{acks}

\clearpage
\section*{GenAI Usage Disclosure}

Beyond their role within the proposed methodology, the GPT-5 and Claude 4 model families were used solely for clarity improvements in writing and functional suggestions in code. All results were reviewed, verified, and finalized by the authors.



\bibliographystyle{ACM-Reference-Format}
\bibliography{custom}

\end{document}